\documentclass[10pt]{article}

\usepackage{booktabs}
\usepackage{censor}
\usepackage{booktabs}
\usepackage{tabularx}
\usepackage{amsmath}
\usepackage{cleveref}
\usepackage{todonotes}
\usepackage{inconsolata}

\usepackage{xcolor}

\crefformat{section}{\S#2#1#3}
\crefformat{subsection}{\S#2#1#3}
\crefformat{subsubsection}{\S#2#1#3}

\usepackage[letterpaper]{geometry}
\usepackage{hicss}
\usepackage{times}
\usepackage[none]{hyphenat}
\usepackage{url}
\usepackage{latexsym}
\usepackage{minted}
\usepackage{indentfirst}
\usepackage{graphicx}
\graphicspath{{images/}}
\usepackage[
    style=apa,
  ]{biblatex}
\title{Sustaining Research Software: A Fitness Function Approach}

\author{Philipp Zech\\
  Department of Computer Science\\University of Innsbruck\\
  {\underline{philipp.zech@uibk.ac.at}} \\\And
  Irdin Pekaric \\
  Department of Computer Science and Information Systems\\University of Liechtenstein\\
  {\underline{irdin.pekaric@uni.li}} \\
}

\date{}

\begin{document}
\maketitle
\begin{abstract}
  The long-term sustainability of research software is a critical challenge, as it usually 
  suffers from poor maintainability, lack of adaptability, and eventual obsolescence. This 
  paper proposes a novel approach to addressing this issue by leveraging the concept of 
  fitness functions from evolutionary architecture. Fitness functions are automated, 
  continuously evaluated metrics designed to ensure that software systems meet desired 
  non-functional, architectural qualities over time. We define a set of fitness functions 
  tailored to the unique requirements of research software, focusing on findability, 
  accessibility, interoperability and reusability (FAIR). These fitness functions act as 
  proactive safeguards, promoting practices such as modular design, comprehensive documentation, 
  version control, and compatibility with evolving technological ecosystems. By integrating 
  these metrics into the development life cycle, we aim to foster a culture of sustainability 
  within the research community. Case studies and experimental results demonstrate the potential 
  of this approach to enhance the long-term FAIR of research software, bridging the gap between 
  ephemeral project-based development and enduring scientific impact.
\end{abstract}

\subsubsection*{Keywords:}

Research, Sustainability, Reproducibility, Repository, Resources, FAIR

\section{Introduction}
\label{sec:introduction}

Research software represents a crucial aspect of academic research. Scientists develop artifacts and 
tools that are often the final product of their research efforts. This software can be reused or evaluated 
by other academics. On the one hand, the reason behind this could be to assess whether their algorithm or
a model performs better against state-of-the-art work. On the other hand, software artifacts could be 
used as a part of the newly implemented pipelines to benefit a better good.

However, despite their importance, there are significant challenges regarding the visibility and 
accessibility of these software artifacts. Recent findings \parencite{pekaric2025we} highlight that a mere 
3\% of papers presented at the Hawaii International Conference on System Sciences (HICSS) from 2017 to 
2024 have functional and publicly available repository containing supplementary materials. This meager 
statistic underscores a broader issue within the research community --- many valuable artifacts remain 
unshared, limiting their reuse and the potential for subsequent advancements in related fields. As \textcite{pekaric2025we} indicate, the lack of shared resources can diminish the transparency 
and reproducibility of research outputs, which are essential for validating findings and fostering 
collaborative progress within academia.

Given the fundamental challenges faced in maintaining and utilizing research software, a strategic shift 
towards the inclusion of fitness functions emerges as an essential solution. These fitness functions serve 
as guiding metrics, ensuring that software fulfills research and adapts to the dynamic landscape of scientific 
research. By utilizing these functions in their artifacts, scientists can promote a culture of sustainable 
practices, enhancing the software’s longevity and compatibility with emerging technologies. This transition 
not only facilitates compliance with FAIR principles --- ensuring that research artifacts are findable, 
accessible, interoperable, and reusable --- but also encourages a collaborative environment where knowledge 
can be efficiently shared and built upon.

In this regard, we propose four fitness functions for sustainable research software. Our work follows the 
Design Science Research (DSR)~\parencite{Wieringa2014DesignSM} paradigm and produces a prototype as an artifact. 
The development of our artifact follows a systematic process, starting with requirements engineering and ending 
with implementing a prototype and its evaluation using a case study. Our artifact is implemented as a solution 
to the following design science problem, outlined using the DSR template~\parencite{Wieringa2014DesignSM}:
\begin{table}[H]
  \small
	\begin{tabularx}{\columnwidth}{lX}
	\toprule
	\textbf{Improve} \textit{research software sustainability} (context) \\
	\textbf{by designing} \emph{a set of fitness functions} (artifact) \\
	\textbf{that satisfy} \emph{research software sustainability} (requirement) \\
	\textbf{to deliver} \emph{FAIR.} (goal) \\
	\bottomrule
  \end{tabularx}
\end{table}
DSR usually refers to an \emph{artifact} as a prototype at Technology Readiness Level 3 
(TRL3), representing a conceptual solution at an early stage of technology development. 
Using a prototype with case-study-based evaluation, our proposal achieves TRL6 (cf.~Sec.~\ref{sec:results}). 
Our artifact is available for download and experimentation from \url{https://github.com/irdin-pekaric/fitness-functions-reusability/}.

\section{Background and Related Work}
\label{sec:background}

The term research software is used to refer to all software applications developed 
by scientists and engineers in an academic setting like universities and research 
centers. This encompasses a wide range of software products, from small throwaway 
scripts developed by one person to huge applications maintained for decades by multiple
research teams around the globe. The developers of these software products are
mainly scientists and domain experts without a formal software engineering education.
For this and other reasons (see below) software 
developed in these domains often lack modern software engineering best practices~\parencite{heaton2015claims}. 
\textcite{matthews2022ex} and \textcite{nowogrodzki2019tips} highlight the
growing importance of software in academia and reveal quality problems of software
developed by researchers. \textcite{noor2022improving}, and \textcite{johanson2018software} 
along with \textcite{heaton2015claims} 
further pinpoint specific issues in scientific software development. They not only 
emphasize the importance of applying software engineering best practices but also provide 
concrete solutions for enhancing the overall quality of software within this field. In 
many scientific software products, the developers, end-users, and stakeholders are the 
same group of scientists and they may be inclined to sacrifice software quality if it 
means they are able to faster reach a working solution~\parencite{lawlor2015engineering}. 
Scientists writing open source scientific software often lack training in software 
engineering and research groups have limited resources to spend on software quality 
assurance~\parencite{georgeson2019bionitio}. This low quality often leads to issues in 
maintainability, performance and usability, which in turn causes difficulties 
in applying, and replicating research.

The substantial involvement of scientists, rather than software engineers, in the 
development of scientific software is rooted in a fundamental disparity. While most
software developers may have some grasp of the functionalities expected from a hotel
booking or accounting software, they are very unlikely to have intuitions as to what
is to be required from software aimed at scientists. In this field, the development process 
is critically dependent on deep domain knowledge. It is often very difficult to gain 
sufficient domain knowledge to tackle scientific applications, as some problems may 
even need PhD level knowledge just to be understood. Hence, in numerous instances, it 
is considered more practical for scientists to bridge the gap in their software engineering 
skills required for a specific project, rather than expecting a software engineer to amass 
the requisite domain knowledge~\parencite{johanson2018software}.

Another distinguishing factor in scientific software development concerns software
requirements. Developers of scientific software often develop software with the aim
of enhancing their understanding of their research field. In such cases, creating
detailed upfront requirement specifications is a challenging task because the requirements
gradually evolve alongside the software development and the increasing
domain knowledge.

\subsection{Fitness Functions}

\begin{figure}[bth]
    \centering
    \includegraphics[width=\columnwidth]{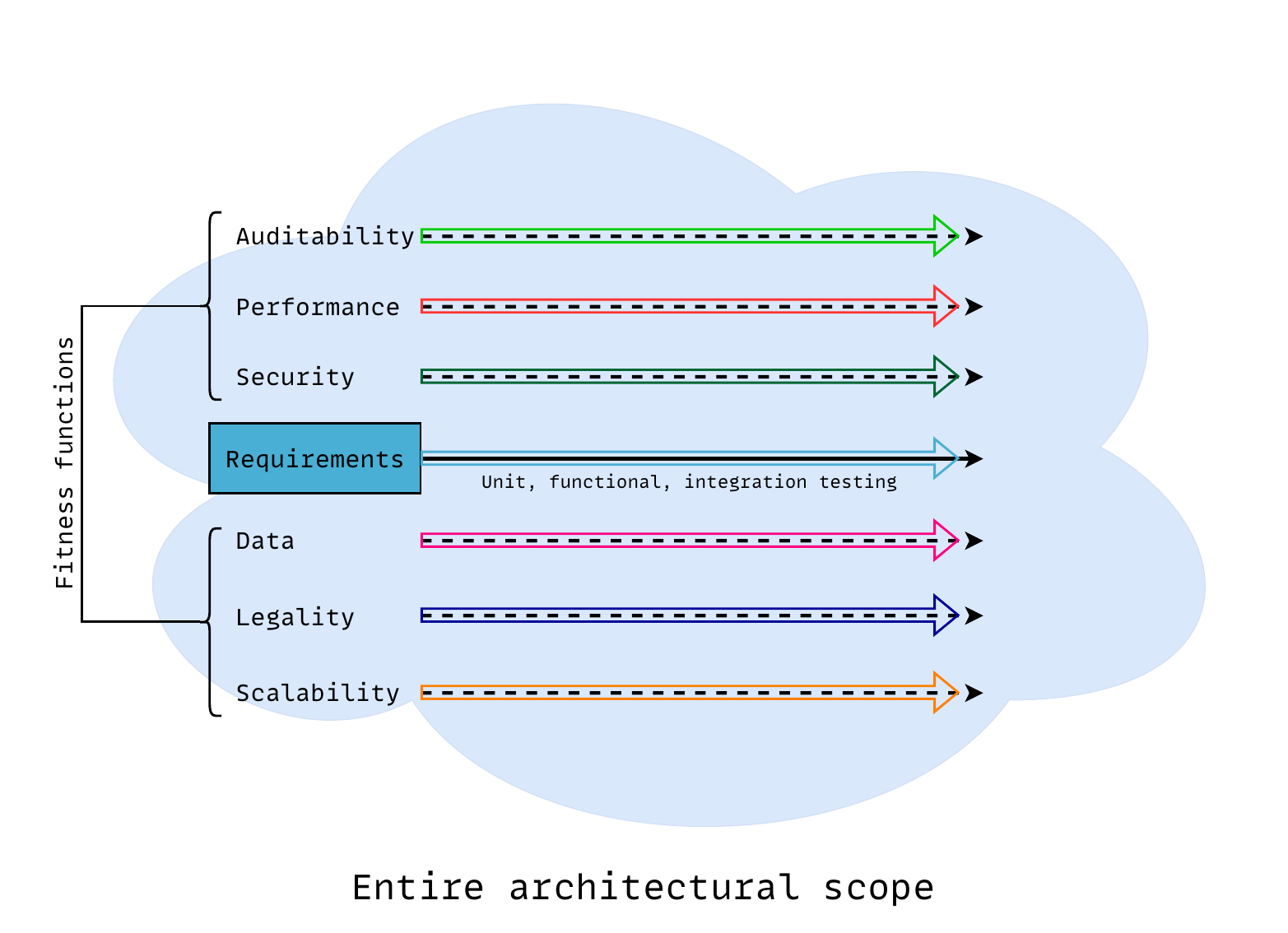}
    \caption{A system's architecture comprises both, requirements and other dimensions which are 
        protected by fitness functions~\parencite{ford2022building}.}
    \label{fig:fitness}
\end{figure}

In \textit{Building Evolutionary Architectures}, \textcite{ford2022building} present 
a methodology for the design of software systems that can \textbf{accommodate change} over time. 
In contrast to conventional architectures that deteriorate with changing requirements, evolutionary 
architectures incorporate fitness functions -- automated checks that uphold essential system 
attributes such as performance, security, and durability (cf.~Fig.~\ref{fig:fitness}). This method 
facilitates \textbf{incremental modification} without compromising the stability of the entire system, 
hence providing sustained maintainability.

An explicit illustration of evolutionary architecture is microservices and continuous deployment. In 
cloud-based systems, autonomous services develop at varying paces, necessitating fitness functions 
that assess service latency to avert performance decline. In Digital Twins, it is essential 
to maintain synchronization between the physical and virtual models. A fitness function for data 
accuracy guarantees that updates adhere to permissible time gaps, facilitating real-time applications 
in smart cities and industrial automation. In the automobile sector, evolutionary concepts facilitate 
the management of over-the-air software upgrades, wherein a fault-tolerance fitness function mitigates 
system faults during the deployment of new vehicle firmware.  

Integrating \textit{fitness functions} into architecture enhances the resilience, scalability, and 
adaptability of software systems. This concept corresponds with self-adaptive systems, AI-driven 
optimization, and model-driven engineering, rendering it a potent strategy for ensuring the longevity 
of intricate software architectures.

We propose that this aforesaid concept of \emph{fitness functions} provides a robust and flexible 
framework for integrating FAIR principles into research software development. By establishing measurable 
restrictions that consistently assess adherence to FAIR principles, fitness functions can actively 
promote ethical and repeatable research methodologies. This not only preserves data integrity and 
computational reliability but also promotes trustworthy and sustainable software ecosystems that 
progress in tandem with scientific advancements.

\subsection{Related Work}

Research software is essential for modern scholarship but frequently encounters sustainability 
issues stemming from poor management methods~\parencite{carver2022survey}. Innovations such as BoneJ2 
illustrate endeavors to enhance software durability and usability through the application of 
optimal programming methods~\parencite{domander2021bonej2}. Recently, the FAIR principles are becoming 
increasingly pertinent to research software, advocating for both compliance and the necessity of open-source 
availability~\parencite{lamprecht2020towards,hasselbring2020fair}. Research Software Science (RSS) promotes 
sustainable software development practices that facilitate scientific discovery~\parencite{heroux2022research}. 
This changing environment requires improved financing mechanisms and acknowledgment for developers to 
create a setting that promotes creativity and reproducibility in research~\parencite{carver2022survey}. 

Despite some initial and early efforts, the current state of sustainability on research software is 
still in an unsatisfactory state. Our proposal (cf.~\cref{ssec:fitness} provides a first systematic 
step to ensure FAIR principles for research software artifacts.

\section{Methodology}
\label{sec:methodology}

Following the practices of DSR~\parencite{Wieringa2014DesignSM} and in light of our 
earlier defined design science problem (cf.~\cref{sec:introduction}, we next 
continue with eliciting relevant requirements for our fitness function framework. 
Observe that these are aligned with the FAIR principles as discussed next.

With relevant requirements established, we continue with presenting our fitness 
function framework for sustainable research software in~\cref{ssec:fitness}.

\subsection{Requirements for Sustainable Research Software}
\label{ssec:reqs}

The FAIR principles, introduced in 2016 by \textcite{wilkinson2016fair}, represent a fundamental 
framework in the domain of scientific data management. The acronym FAIR represents four key principles: Findable, 
Accessible, Interoperable, and Reusable. While these principles were initially conceptualized for scientific data, 
their application has been later extended to also cover research software. The application of FAIR principles to 
research software addresses the critical need for improved discoverability, accessibility, and reusability of 
scientific tools and resources. These principles ensure that both data and the software used to analyze it can 
be effectively shared, integrated, and built upon by the broader scientific community, which fosters a more 
collaborative, efficient, and transparent scientific ecosystem. We describe each aforementioned principle below.

\noindent \textit{Findability.} Ensuring that research software can be readily discovered by both humans and 
machine systems is crucial for effective utilization of research software in the scientific community \parencite{weigel2020making}. 
For example, this can be achieved through the assignment of unique and persistent identifiers to software 
artifacts, such as Digital Object Identifiers (DOIs). Additionally, findability implies the provision of rich 
metadata that comprehensively describes the software, as well as the registration or indexing of the software 
in searchable resources such as software repositories (e.g., GitHub, GitLab), academic databases (e.g., Zenodo, 
figshare), or domain-specific catalogs.

\noindent \textit{Accessibility.} Software should be obtainable through standardized protocols, utilizing open, 
free, and universally implementable access methods~\parencite{lamprecht2020towards}. This includes version control 
systems (e.g., Git) or package managers (e.g., pip for Python, npm for JavaScript), which ensures that researchers 
from diverse technical (and other) backgrounds can easily obtain and deploy the software. Moreover, this principle 
also addresses the aspect of longevity, implying that the software and metadata should be preserved even after the 
active development phase has concluded.

\noindent \textit{Interoperability.} Software needs to possess the capability to function seamlessly with other 
software, systems and data. This is due to the fact that complex scientific questions often require the integration 
of multiple tools and datasets. Interoperability can be achieved by using formal, accessible, shared, and broadly 
applicable languages for knowledge representation~\parencite{hazra2021comprehensive} such as standard data formats 
(e.g., CSV, JSON, HDF5), well-defined APIs (Application Programming Interfaces) and widely accepted communication 
protocols (e.g., REST, GraphQL). These standards ensure exchange of data and functionalities without the need for 
custom integration work for each new combination of tools.

\noindent \textit{Reusability.} Maximizing the value of research software by enabling its application in diverse 
contexts aims to accelerate scientific progress and foster innovation. This involves several aspects including the 
clear specification of usage licenses (preferably open-source licenses as indicated by \textcite{makitalo2020opportunistic}), 
provision of detailed provenance information~\parencite{capilla2019opportunities} (e.g., software’s development history, 
contributors, and any modifications or extensions),  adherence to domain-relevant community standards and the 
need for comprehensive documentation~\parencite{saied2018improving} (e.g., user guides, API references and examples 
of use cases). Lastly, the reusability aspect also benefits from active community engagement~\parencite{katz2018community}, 
which often includes mechanisms for user feedback, bug reporting, and feature requests. 

\begin{figure}[bth]
    \centering
    \includegraphics[width=\columnwidth]{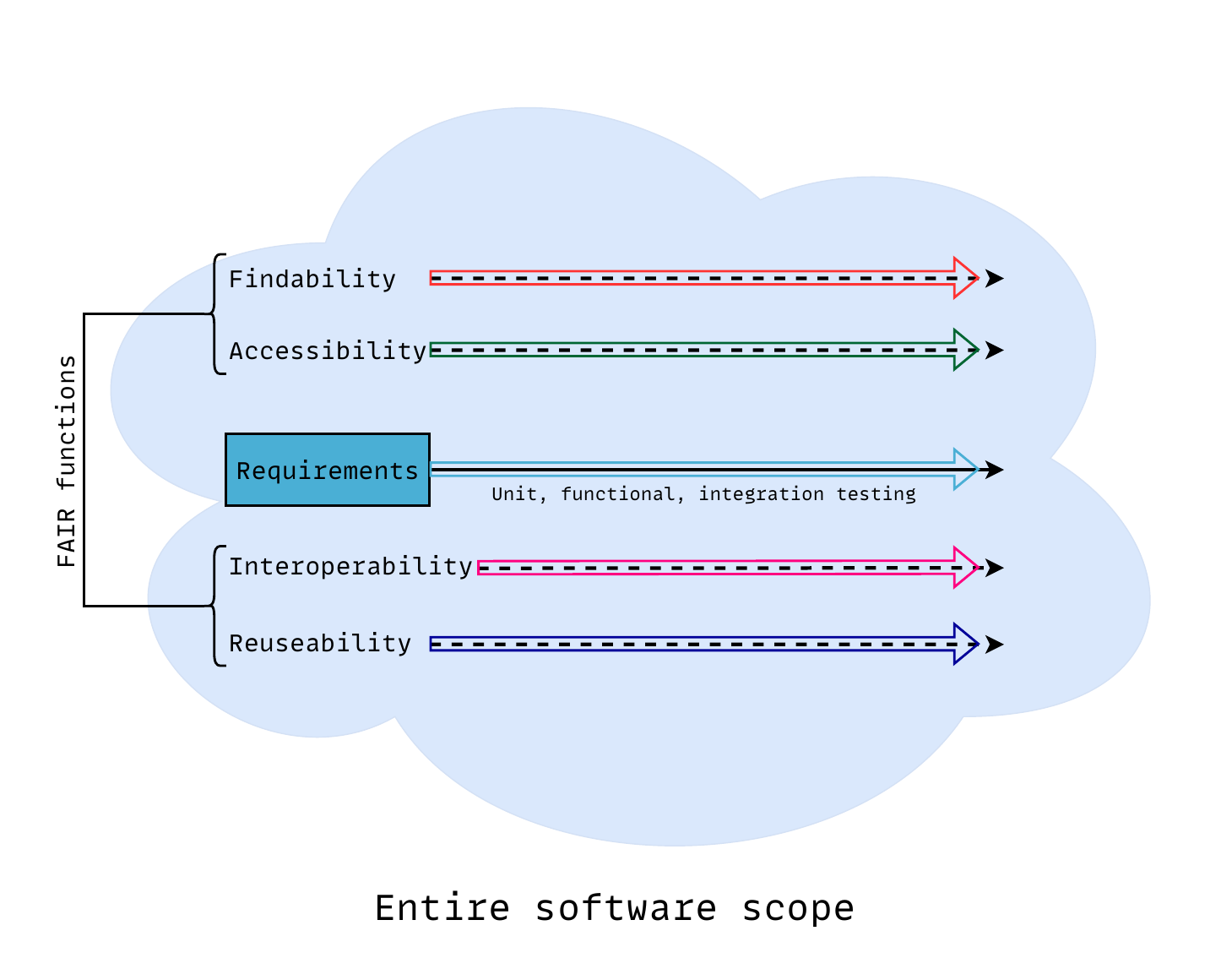}
    \caption{Research software comprises both, requirements and FAIR dimensions for sustainable 
        research software which are protected by fitness (cf.~FAIR) functions.}
    \label{fig:fitness-fair}
\end{figure}

Fig.~\ref{fig:fitness-fair} aligns the FAIR dimensions with our proposal of employing fitness functions to 
ensure FAIR principles.

\subsection{Fitness Functions for Sustainable Research Software}
\label{ssec:fitness}

Given the requirements as elicited in \cref{ssec:reqs}, in the following we provide an 
abstract, implementation-independent definition of fitness functions for sustainable research 
software based on set theory notation~\parencite{hausdorff2021set}. All fitness functions are defined 
as \emph{atomic} fitness functions, i.e., they run in a singular context and exercise one particular 
aspect~\parencite{ford2022building}.

\subsubsection{Findability}

Findability ensures that a research artifact is uniquely identifiable and can be reliably 
located. Accordingly, we define a fitness function that verifies whether the artifact is 
discoverable. If the artifact is found, $findability$ returns its \text{Location} (e.g., a URI); 
otherwise, it returns null $\emptyset$ (cf.~null). The formal definition of $findability$ is as 
follows:
\[
\text{findability}: \text{Metadata} \to \{ \text{Location}, \emptyset \}
\]
\begin{tabular}{@{}p{2.5cm}p{5cm}@{}}
\toprule
\textbf{Precondition} & \( \forall m \in \text{Metadata}, m \neq \emptyset \) (The input metadata must be valid and not null). \\
\textbf{Postcondition (found)} & If the artifact is found, the function returns a valid location: \( \text{findability}(m) = l, \, l \in \text{Location} \). \\
\textbf{Postcondition (not found)} & If the artifact is not found, the function returns null: \( \text{findability}(m) = \emptyset \). \\ \bottomrule
\end{tabular}

\subsubsection{Accessibility}

Complementary to findability, accessibility ensures that a research artifact -- once its location 
is known -- can be retrieved using standard protocols (e.g., HTTP or FTP). This typically includes 
not only the artifact itself but also its associated data and documentation. Accordingly, if an 
artifact is accessible, $accessibility$ returns a $Artifcact$ (i.e., the downloaded data) for further 
use; otherwise, it returns null $\emptyset$ (cf.~null). The formal definition of $accessibility$ 
is as follows:
\[
\text{accessibility}: \text{Location} \to \{ \text{Artifact}, \emptyset \}
\]
\begin{tabular}{@{}p{2.5cm}p{5cm}@{}}
\toprule
\textbf{Precondition} & \( l \neq \emptyset \) (The input location must valid and not be null). \\
\textbf{Postcondition (found)} & If the location is accessible, the function returns the Artifact (the downloaded data): \( \text{accessibility}(l) = b, \, b \neq \emptyset \). \\
\textbf{Postcondition (not found)} & If the location is not accessible, the function returns null: \( \text{accessibility}(l) = \emptyset \). \\ \bottomrule
\end{tabular}

\subsubsection{Interoperability}

Interoperability ensures that a research artifact (software or data) can function correctly within 
a given environment, including integration with other tools, data formats, and systems. This usually 
comprises
\begin{enumerate}
    \item the technical environment, i.e., the hardware and software ecosystem, such as
        operating system and platform, programming languages and runtime dependencies, or software 
        libraries and frameworks; and
    \item the data environment, i.e., formats, structures, and standards, such as data formats 
        and file structures, metadata standards and ontologies, or APIs and data access protocols.
\end{enumerate}
As such, actual implementations of $interoperability$ may vary depending on the context (cf.~\cref{sec:results}). 
Given that for a specific environment the artifact can interact in and with it, $interoperability$ returns $True$; 
otherwise, it returns $False$. The formal definition of $ineroperability$ is as follows:
\[
\text{interoperability}: (\text{Artifact}, \text{Environment}) \to \{ \text{True}, \text{False} \}
\]
\begin{tabular}{@{}p{2.5cm}p{5cm}@{}}
\toprule
\textbf{Precondition} & \( \forall e \in \text{Environment}, e \neq \emptyset, a \neq \emptyset \) (The artifact and environment are not null). \\
\textbf{Postcondition (True)} & If the artifact is able to interact correctly with the given environment (e.g., adhering to standard data formats, 
    using common APIs, or following compatible protocols), the function returns success: \( \text{interoperability}(a, e) = \text{True} \). \\
\textbf{Postcondition (False)} & If the artifact fails to integrate properly within the environment (e.g., incompatible APIs, unsupported data 
    formats), the function returns failure: \( \text{interoperability}(a, e) = \text{False} \). \\ \bottomrule
\end{tabular}

\subsubsection{Reusability}

Reusability ensures that a research software artifact can be efficiently adapted and applied in different 
contexts beyond its original purpose, with minimal modification. As such -- analogously to interoperability --  
it also comprises a composition of aspects, viz:
\begin{enumerate}
    \item technical resuability such as a modular and extensive codebase, versioning and compatibility, 
        or software packaging and distribution; 
    \item documentation and usability such as a usage tutorials and comprehensive examples, or 
        ease of installation and deployment; and
    \item licensing and governance such as applied licensing and distribution model or community 
        support.
\end{enumerate}
Given these quite conceptual interpretation of reusability (cf.~\cref{ssec:reqs}), implementations of 
actual fitness functions targeting assessing reusability are very much domain-, artifact- and application-specific 
(cf.~\cref{sec:results}). In addition, as to the diverse nature of targeted aspects (e.g., well-documented, 
openly licensed, actively maintained, or technically easy to integrate), given the actual 
implementation language and employed framework, not all aspects (e.g., packaging and distribution in 
the event of C++-based artifacts is quite rudimentary) may be assessed. Given that an artifact fulfills 
encoded  aspects, $reusability$ returns $Success$; otherwise, it returns $False$. The formal definition 
of $reusability$ is as follows:
\[
\text{reusability}: \text{Artifact} \to \{ \text{True}, \text{False} \}
\]
\begin{tabular}{@{}p{2.5cm}p{5cm}@{}}
\toprule
\textbf{Precondition} & \( a \neq \emptyset \) (The artifact is not null). \\
\textbf{Postcondition (True)} & If the assessed aspects are fulfilled, e.g., the artifacts is well-documented, 
    openly licensed, actively maintained, or technically easy to integrate, the function returns success: \( \text{resusability}(a) = \text{True} \). \\ 
\textbf{Postcondition (False)} & If any of the assessed aspects is not fulfilled, reuse is effectively impossible 
    and the function returns failure: \( \text{resusability}(a) = \text{False} \). \\ \bottomrule
\end{tabular}

\vspace{.5cm}
This section has formally introduced a set of fitness functions that create a methodical and quantifiable
method for guaranteeing the long-term sustainability of research software in accordance with the FAIR principles. 
Establishing formal criteria for findability, accessibility, interoperability, and reusability facilitates automated 
evaluations that consistently confirm if software artifacts are discoverable, retrievable, compatible with other 
systems, and adaptable for future utilization. This approach facilitates the direct integration of sustainability 
into the research software development life cycle, enabling researchers to proactively detect and rectify deficiencies 
in their artifact's maintainability and usability. Our proposal not only enhances software quality but also promotes 
more trustworthy, transparent, and collaborative scientific environments. Integrating these fitness capabilities into 
research software workflows enables academics to promote standardized best practices, guarantee the long-term usability 
of research artifacts and results, and mitigate the danger of obsolescence. 

The last phase of this work now comprises converting these formal definitions into tangible implementations, 
illustrating the practical application of these fitness functions to effectively improve the sustainability of 
research software.

\section{Results}
\label{sec:results}

In the following we report on results achieved when implementing our proposed 
framework for \emph{Pekaric and Apruzzese: We Provide our Resources in a Dedicated Repository}~\parencite{pekaric2025we}. 
Pekaric and Apruzzese's artifact contains scripts, classification sheets, and findings from analyzing the 
transparency and availability of supplementary materials linked to papers accepted at HICSS. It is provided 
to promote transparency in research practices, enhance reproducibility and support the reuse of scientific 
artifacts, which makes it a perfect artifact for evaluation of the proposed approach. Our resources and source 
code are available via \url{https://github.com/irdin-pekaric/fitness-functions-reusability/}. In the following, 
we discuss the FAIR fitness functions implemented for Pekaric and Apruzzese's artifact.

\subsection{FAIR Functions}

To move from abstract principles to actionable practices, we developed a suite of four targeted fitness 
functions—each directly aligned with one of the FAIR dimensions. These functions are implemented as lightweight, 
modular scripts, each designed to assess a single property of a research software artifact. By capturing concrete, 
measurable criteria, the functions serve as practical checkpoints that make it possible to evaluate the FAIR compliance 
of software in a consistent and repeatable way. Our goal was not merely to propose FAIR as an ideal, but to create 
tools that allow researchers to verify, demonstrate, and maintain it over time.

The functions and the checks they implement are ``Git-aware'' and can be extended to support GitLab, Bitbucket or 
institutional Git services. This makes it especially useful for large-scale audits of research software portfolios, 
grant compliance checks or onboarding validations in collaborative projects.

The \texttt{findability} function addresses a common shortfall in research software: poor discoverability. To 
evaluate whether an artifact can be reliably located, this function requires a set of resolvable links, DOIs, 
or repository identifiers as input. If a valid location is found and confirmed to exist, the function returns 
the location; if not, it flags the artifact as undiscoverable. In practice, this check ensures that research 
outputs can be referenced, cited, and indexed—making them visible not just to humans, but also to automated 
systems that support scholarly communication.

\begin{figure}[htbp]
    \centering
    \includegraphics[width=\columnwidth]{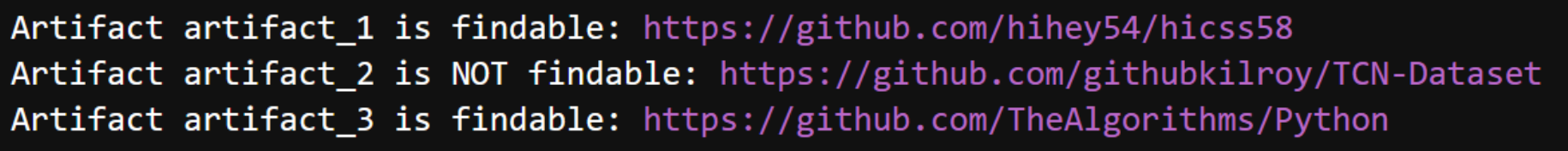}
    \caption{Output of the \texttt{FindabilityChecker}.}
    \label{fig:find}
\end{figure}

Our \texttt{FindabilityChecker} evaluates whether a software artifact’s 
Uniform Resource Identifier (URI) is reachable via the web (\textit{https://github.com/hihey54/hicss58}). It 
performs a HTTP HEAD request to test the resolvability of the URI provided in the artifact metadata. A successful 
response confirms that the artifact's URI is valid and that it is actually reachable at the time of the check 
as shown in Figure \ref{fig:find}\footnote{The remaining two artifacts that are present in Figure \ref{fig:find} serve as illustrative counterexamples, highlighting cases where the associated repositories could not be located. These are also present in the output of other checkers to demonstrate the full potential of our tool.}.

The simplicity of the check conceals its significance. In many real-world scenarios, datasets and codebases are 
referenced with links that may decay over time (``link rot''). The \texttt{FindabilityChecker} provides a fast 
and scalable way to catch such issues. It is well-suited for integration into periodic validation workflows for 
various repositories, DOI registries or research software archives. This allows maintainers to track and resolve 
broken references before they impact reproducibility or accessibility.

Building on this, the \texttt{accessibility} function verifies whether the software is actually retrievable 
from its declared location. This involves checking that the repository or archive is accessible over standard 
web protocols, and can be downloaded. A successful check downloads the artifact, while failure may indicate 
various issues such as insufficient permissions. This function is essential to ensure that published research artifacts 
remain accessible, regardless of project or institutional timelines.

\begin{figure}[htbp]
    \centering
    \includegraphics[width=\columnwidth]{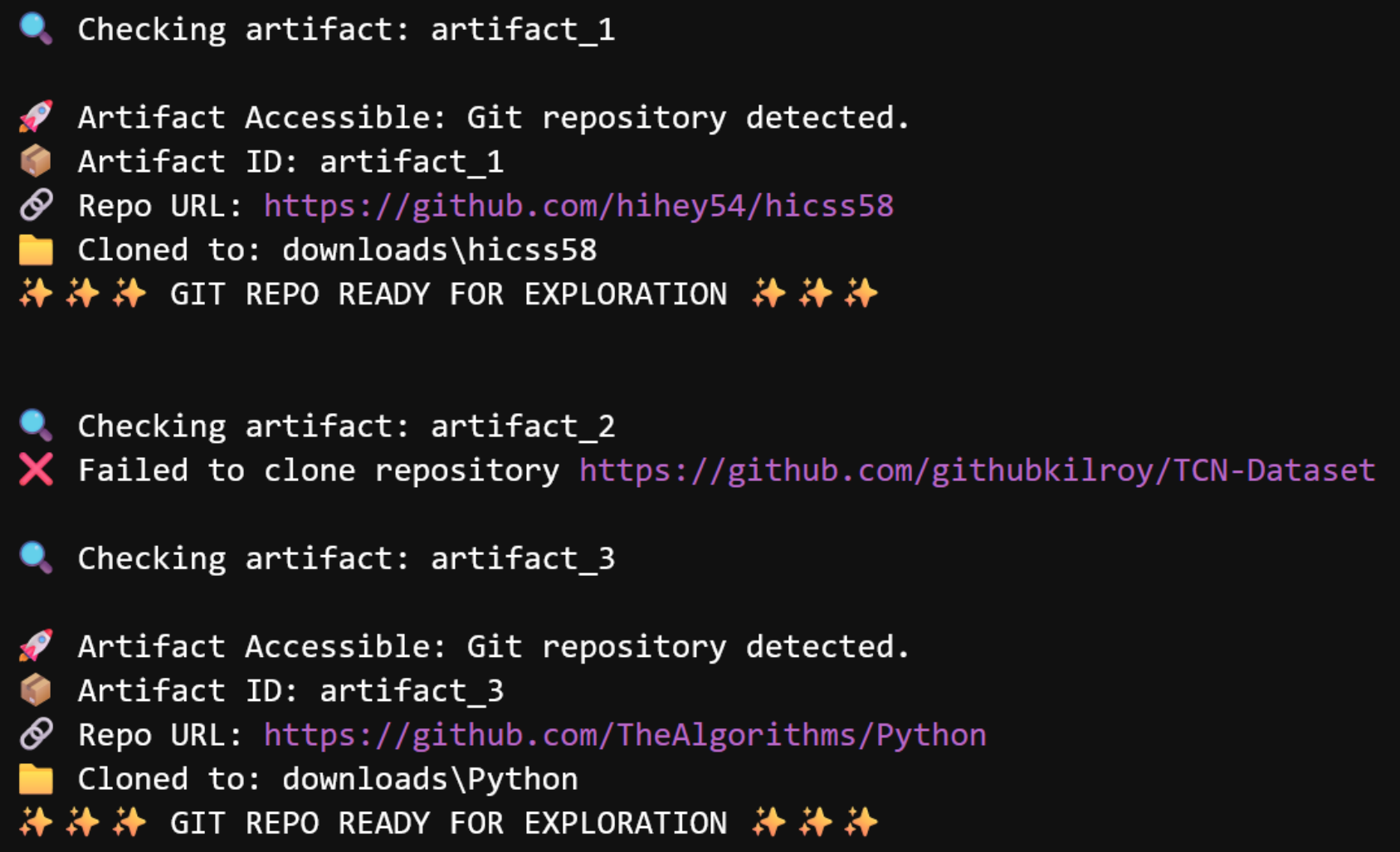}
    \caption{Output of the \texttt{AccessibilityChecker}.}
    \label{fig:access}
\end{figure}

Following the findability test, the \texttt{AccessibilityChecker} attempts to retrieve the content from the 
identified URI. In our use case, the checker issues a GET request and successfully downloads the artifact. 
This verifies not only that the link is ``live'', but that it also leads to a usable digital object. Our 
\texttt{AccessibilityChecker} (see Figure \ref{fig:access}) then saves the artifact locally using the existing 
name of the repository (``hicss58'').

This step reflects a deeper layer of the FAIR principles: accessibility is not guaranteed by findability alone. 
Some links may resolve but lead to login walls or corrupted files. The checker handles these edge cases and 
reports artifacts that are findable but not retrievable. This distinction is especially relevant for automated 
meta-research tools, which must separate technical failures from policy-imposed access restrictions (e.g., 
licensing or downtime due to a ban).

The \texttt{interoperability} function examines whether the software can operate within a defined computational 
environment. For our Python-based prototype, this means verifying the presence of an interpreter and a complete 
\texttt{requirements.txt} file, and testing whether dependencies can be installed without conflict. If the environment 
can be reliably reproduced and the software executes as intended, the artifact is deemed interoperable. This 
function simulates what a future user --- or a replication study --- would experience when trying to integrate the 
software into a larger pipeline or research workflow.

\begin{figure}[htbp]
    \centering
    \includegraphics[width=\columnwidth]{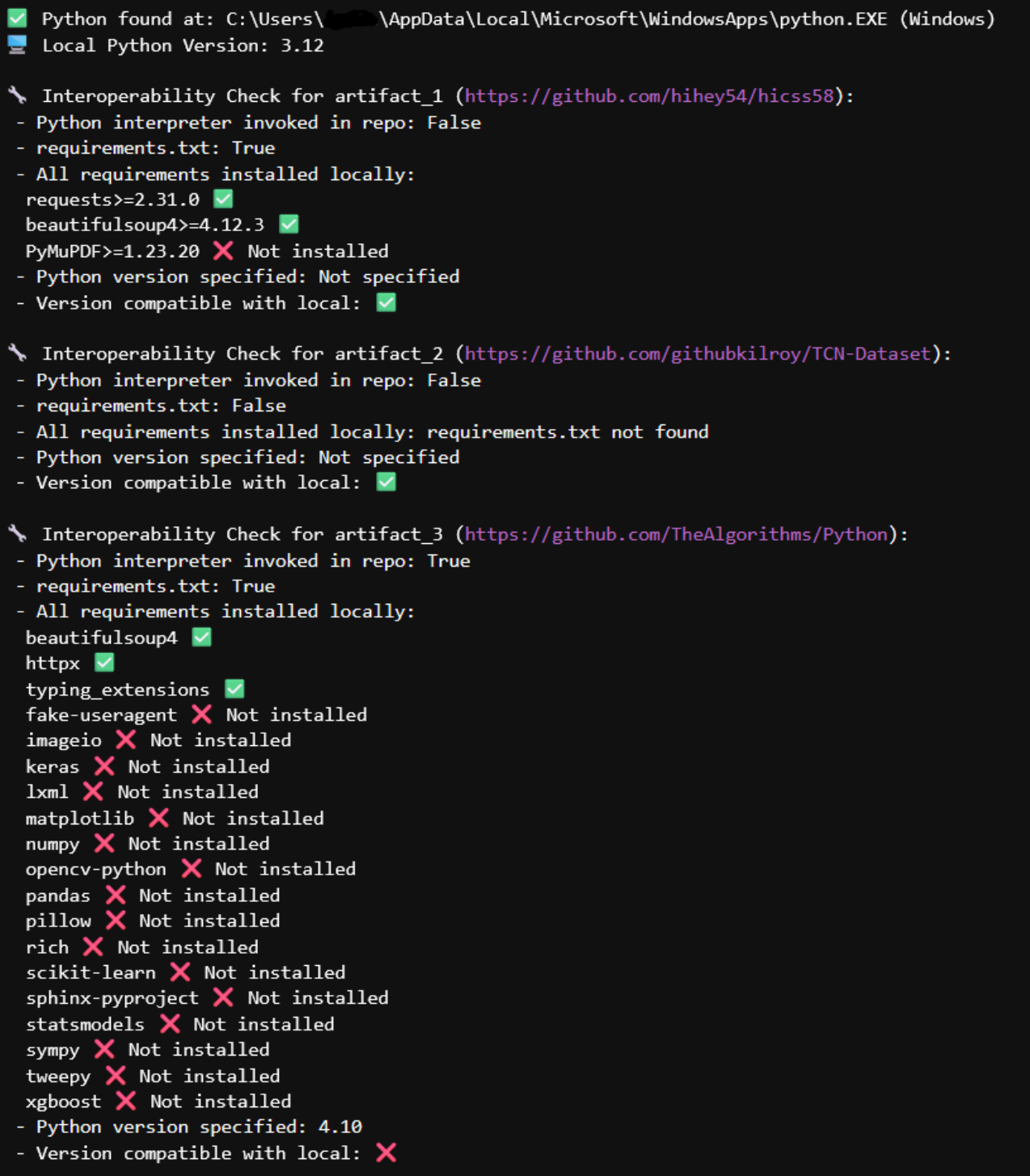}
    \caption{Output of the \texttt{InteroperabilityChecker}.}
    \label{fig:intero}
\end{figure}

Our \texttt{InteroperabilityChecker} performs a multifaceted evaluation of the software artifact’s readiness 
in this regard. It inspects the GitHub repository linked to the artifact (\textit{https://github.com/hihey54/hicss58}) 
and checks for several signs of good software hygiene and platform-neutral execution. In our use case, the 
checker verifies the existence of a \texttt{requirements.txt file}, CI/CD workflows invoking a Python interpreter, 
and metadata files such as \texttt{setup.py} or \texttt{pyproject.toml}. Furthermore, it compares the required 
Python version (if specified) against the Python version installed locally, ensuring environmental compatibility. 
It also attempts to resolve and validate all listed dependencies using \texttt{pkg\_resources}, allowing us to 
detect undeclared or broken requirements. The output of the checker is shown in Figure \ref{fig:intero}. 

This procedure simulates what a developer or automated agent would encounter when trying to run the artifact. 
The \texttt{InteroperabilityChecker} flags whether the artifact is ready to be integrated into workflows, 
continuous integration environments or larger modular pipelines. Importantly, the tool is structured to allow 
domain-specific extensions—e.g., Java \texttt{.pom} and \texttt{.jar} inspections, or R’s DESCRIPTION files—enabling 
broader ecosystem support.

Finally, the \texttt{reusability} function assesses whether the software is prepared for use beyond its original 
context. This includes checking for the presence of a license that permits reuse, a \texttt{README.md} file that 
provides installation and usage guidance, and other indicators of maintainability such as modular code structure 
or versioning (depending on an artifacts size). Where present and well-formed, these elements mark the software 
as reusable. In their absence, even technically sound artifacts risk becoming obscure or unusable. This function 
therefore emphasizes the human-facing aspects of sustainability: clarity, legality and adaptability.

\begin{figure}[htbp]
    \centering
    \includegraphics[width=\columnwidth]{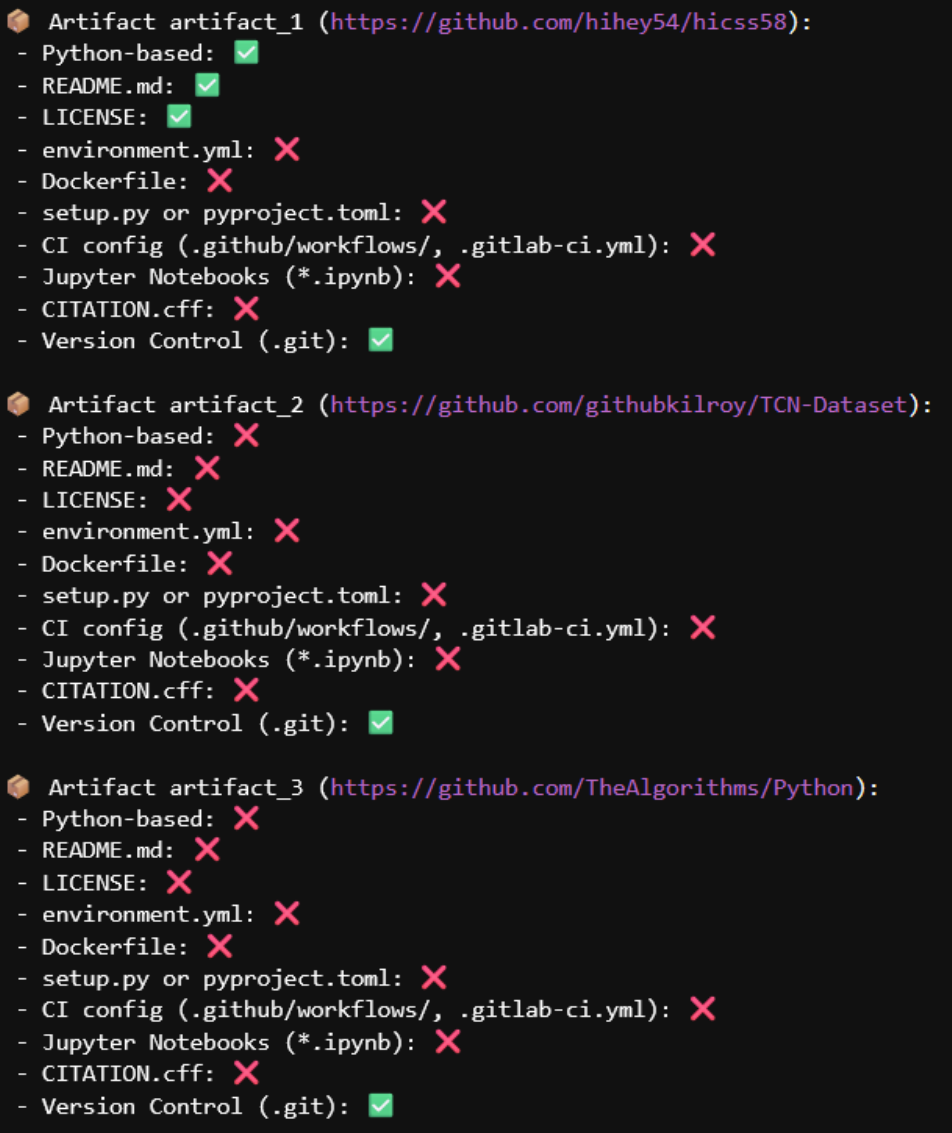}
    \caption{Output of the \texttt{ReproducabilityChecker}.}
    \label{fig:repro}
\end{figure}

Our \texttt{ReproducibilityChecker} evaluates whether a repository contains the elements necessary to support 
computational reproducibility. In the selected use case, the checker scans for multiple files and repository 
features that collectively indicate good reproducibility practices. These include a \texttt{README.md} (for usage 
documentation), a \texttt{LICENSE} or \texttt{LICENSE.txt} (for legal reuse), and Dockerfile or \texttt{environment.yml} 
files (for environment encapsulation). The \texttt{ReproducibilityChecker} (see Figure \ref{fig:repro}) also checks 
for the presence of computational notebooks (\texttt{.ipynb}), automated workflows such as GitHub Actions 
(\texttt{.github/workflows/*.yml}), and even academic-specific metadata such as \texttt{CITATION.cff}.

Each of these elements helps to make the artifact usable, verifiable and repeatable by others. For example, 
a Dockerfile helps guarantee that the software behaves the same across machines. A CITATION.cff file increases 
discoverability and promotes proper attribution. Thus, the \texttt{ReproducibilityChecker}  allows researchers 
and software engineers to quickly check a repository’s maturity and reproducibility potential.

Together, these fitness functions form a coherent, low-overhead mechanism for evaluating FAIR compliance in 
research software. By running them periodically, researchers can catch lapses early and maintain high standards 
without imposing additional burdens during development. While our current implementation focuses on the core 
FAIR dimensions, the framework is inherently extensible and can evolve to include aspects such as security, 
provenance, or community engagement. In this way, our work aims to bridge the gap between aspirational principles 
and sustainable, everyday practices in scientific software development. All implementation details are available 
from our artifact via \url{https://github.com/irdin-pekaric/fitness-functions-reusability/}.

\section{Discussion}
\label{sec:discussion}

Our implementation of FAIR-aligned fitness functions demonstrates that sustainability in research software is 
not merely a theoretical goal, but a measurable and actionable property. By making each FAIR dimension testable 
through minimal, focused scripts, we lower the barrier for self-assessment and promote a culture of routine 
quality assurance. Importantly, the checks we developed do not require advanced infrastructure or specialized 
knowledge—they can be executed locally, by researchers themselves, at any stage of a project. This decentralization 
of responsibility is essential in academic environments where centralized QA resources are often unavailable.

One of the key observations from our experimental application is that small, seemingly administrative 
omissions --- such as the absence of a license or the lack of proper dependency specification --- can significantly 
reduce a project’s long-term utility. These are not difficult problems to fix, yet they frequently go unnoticed. 
By codifying FAIR expectations into executable functions, we offer a mechanism that draws attention to these 
issues early, before they become structural barriers to reuse, replication, or maintenance.

While our implementation targets Python-based artifacts, the underlying principles are language-agnostic. 
The methodology can be adapted to any environment with equivalent notions of packaging, metadata, and dependency 
management. The explicit separation of concerns—one function per principle—also provides a template for extension. 
Additional checks, such as those for security, provenance, or even machine-actionable metadata, can be introduced 
as separate modules without affecting the integrity of the core system. This design makes the approach not only 
practical for today’s research needs but also adaptable to future standards and evolving best practices.

Periodic execution of these checks --- either as part of CI pipelines or manual verification routines --- encourages 
a shift from reactive curation (fixing issues when others report them) to proactive stewardship. Over time, 
such practices have the potential to recalibrate community norms, embedding sustainability into the everyday 
development of research software.

Our approach meets the criteria for TRL6 (cf.~\cref{sec:introduction}), having been demonstrated with a 
real-world research software artifact (cf.: \url{https://github.com/irdin-pekaric/fitness-functions-reusability/}) and within 
an operationally relevant academic environment. This situates our framework beyond laboratory validation, 
indicating its readiness for adoption and evaluation in authentic research software workflows~\cite{Wieringa2014DesignSM}.

\subsection{Sustainability Implications}

From a sustainability perspective, the approach we propose has several strengths. First, it encourages forward 
compatibility by enforcing conventions that support long-term maintenance and reuse. Second, it improves 
reproducibility --- not just of scientific results, but of the software environments in which those results are 
generated. Third, it fosters transparency and trust by making research artifacts discoverable, accessible, 
and verifiably documented. All of these outcomes align directly with current priorities in open science and 
funder-mandated data stewardship.

The simplicity of the fitness function model is key to its sustainability impact. By relying on a minimal 
set of measurable, automatable properties, we avoid the pitfalls of overly complex frameworks that may themselves 
become difficult to maintain. Moreover, our model enables researchers to “self-police” FAIR compliance without 
needing to wait for third-party validation. In doing so, it democratizes software sustainability and embeds 
good practices at the grassroots level of research work.

\subsection{Limitations and Future Work}

While our approach offers a practical method for assessing the FAIR compliance of research software, it is 
not without limitations. Most notably, the current implementation remains manually configured and tailored to 
a specific programming ecosystem, e.g., Python. As such, applying the same fitness functions to software 
written in other languages (e.g., C++, R, or Java) would require reimplementation or adaptation of environment-specific 
logic. Additionally, our framework presently focuses on a narrow slice of the broader sustainability landscape: 
the four FAIR dimensions. Other critical aspects, such as security, provenance, community governance, and 
archival persistence, are not yet represented in our checks.

Another limitation stems from the relatively coarse granularity of the fitness functions. While each function 
provides a binary result (pass/fail), this may not sufficiently capture degrees of compliance or allow for 
nuanced feedback in more complex software scenarios. Furthermore, the system does not yet offer a standardized 
reporting mechanism or a way to persist results across project versions or evaluation sessions.

To address these issues, future work will focus on formalizing the development of fitness functions themselves. 
By specifying each function in terms of typed inputs, expected outputs, execution contexts, and side-effects, 
we can treat them as first-class entities that can be composed and validated. We 
also envision the creation of a reusable library or toolkit that allows researchers to instantiate predefined 
fitness functions with minimal configuration. This would enable adoption across a broader spectrum of scientific 
domains and technical stacks, while promoting consistency in implementation.

\section{Conclusion}
\label{sec:conclusion}

In this paper, we introduced a pragmatic and extensible framework for evaluating the sustainability of 
research software through the lens of the FAIR principles. By implementing lightweight, modular fitness 
functions where each targets a distinct dimension of FAIR, we demonstrated how abstract ideals can be 
transformed into operational checks that are automatable, repeatable, and adaptable across projects.

Our approach is intentionally minimal yet impactful: it empowers researchers to assess the health of 
their software artifacts using everyday tools, without requiring deep infrastructure or third-party 
certification. Through practical experimentation, we showed how even basic FAIR checks can reveal 
critical sustainability gaps, such as missing licenses or opaque dependencies, and how addressing 
these gaps early can dramatically improve long-term usability and reproducibility.

At the same time, we acknowledge that the current implementation represents an early-stage prototype, 
with room for refinement. Future work will focus on generalizing the model, building a reusable instantiation 
library, expanding the scope of assessed properties, and integrating orchestration and reporting mechanisms. 
These enhancements will support broader adoption across disciplines and software stacks, and will help embed 
FAIR assessments as a regular part of research software development.

Ultimately, we argue that sustainability must be treated not as a retrospective exercise, but as an 
embedded practice. By making FAIRness verifiable, visible, and actionable, the fitness function approach 
creates a pathway toward more transparent, maintainable, and impactful scientific software. Our hope 
is that this work serves not only as a proof of concept, but as an invitation to the research community 
to adopt, extend, and evolve these practices in service of open and enduring scholarship.

Unlike traditional software quality assessment tools, which often focus on general software engineering 
metrics~\cite{hutcheson2003software}, our framework specifically operationalizes the FAIR principles 
through targeted fitness functions tailored to the unique challenges of research software sustainability. 
This novel approach integrates continuous, principle-driven evaluation into the research life cycle, thereby 
addressing key gaps in reproducibility and long-term usability.

\printbibliography

\end{document}